\documentclass{rspublic}
\usepackage{graphicx,natbib}

\newcommand{\aap}{{\em Astr. \& Ap.}}       
\newcommand{\apj}{{\em Ap. J.}}         
\newcommand{\apjs}{{\em Ap. J. Suppl.}}         
\newcommand{\mnras}{{\em MNRAS}}            
\newcommand{\nat}{{\it Nature}}
\newcommand{\nature}{{\it Nature}}

\newcommand{\meszaros}{M\'{e}sz\'{a}ros}
\newcommand{\peterm}{M\'{e}sz\'{a}ros}

\begin{document}

\title[X-ray Flares]{X-ray Flares in Early GRB Afterglows}

\author[D. N. Burrows and others]{D. N. Burrows$^1$,
  A. Falcone$^1$, G. Chincarini$^{2,3}$, 
 D. Morris$^1$, P. Romano$^2$, J. E. Hill$^{4,5}$, O. Godet$^6$,
 A. Moretti$^2$,
 H. Krimm$^4$, J. P. Osborne$^6$, J. Racusin$^1$, V. Mangano$^7$, K. Page$^6$, M. Perri$^8$, M. Stroh$^1$, and the {\it Swift} XRT team}

\affiliation{$^1$Department of Astronomy \& Astrophysics, The Pennsylvania
  State University, 525 Davey Lab, University Park, PA 16802, USA \\
$^2$INAF-Osservatorio Astronomico di Brera, Via Bianchi 46, 23807
Merate, Italy \\
$^3$Departimento di Fisica, Universit\`{a} degli studi di Milano-Bicocca,
Piazza delle Scienze 3, 20126 Milano, Italy \\
$^4$NASA/Goddard Space Flight Center, Greenbelt, MD 20771, USA \\
$^5$Universities Space Research Association, 10211  
Wincopin Circle, Suite 500, Columbia, MD, 21044-3432, USA \\
$^6$Department of Physics \& Astronomy, University of Leicester,
Leicester LE1 7RH, UK \\
$^7$INAF-Instituto di Astrofisica Spaziale e Fisica Cosmica, Sezione di Palermo, via U. La Malfa 153, I-90146 Palermo, Italy \\
$^8$ASI Science Data Center, via G. Galilei, I-00044 Frascati, Italy}

\label{firstpage}

\maketitle

\begin{abstract}{Gamma-ray Bursts; GRB Afterglows; X-ray Flares}
The {\it Swift} X-ray Telescope (XRT) has discovered that 
flares are quite common in 
early X-ray afterglows of Gamma-Ray Bursts (GRBs), being
observed in roughly 50\% of afterglows with prompt followup observations.  
The flares range in fluence from a few percent to $\sim 100$\% 
of the fluence of the prompt emission (the GRB).  Repetitive flares
are seen, with more than 4 successive flares detected by 
the XRT in some afterglows.  The rise and fall times of the flares are 
typically considerable smaller than the time since the burst.  
These characteristics suggest that the flares are related to the prompt emission 
mechanism, but at lower photon energies.
We conclude that the most likely cause of these flares is late-time
activity of the GRB central engine.
\end{abstract}

\section{Introduction}
The advent of modern Gamma-Ray Burst (GRB) astronomy occurred in 1997 with the
dramatic discovery by the {\it Beppo-SAX} satellite of the first GRB afterglow
\citep{Costa97}.
This led to the identification of GRB host galaxies, the
determination of their redshifts (and hence, their distances), and the
measurement of important physical properties such as total energy.  
These observations provided strong evidence supporting the
fireball model \citep[][and references therein]{Meszaros02}, which attributes the burst itself and the
subsequent afterglow to shocks generated by a
highly relativistic fireball ejected from the GRB progenitor.

\begin{figure}
     \centering
     \parbox{2.4in}{
     \includegraphics[width=2.3in,angle=0,bb=72 230 550 617, clip]{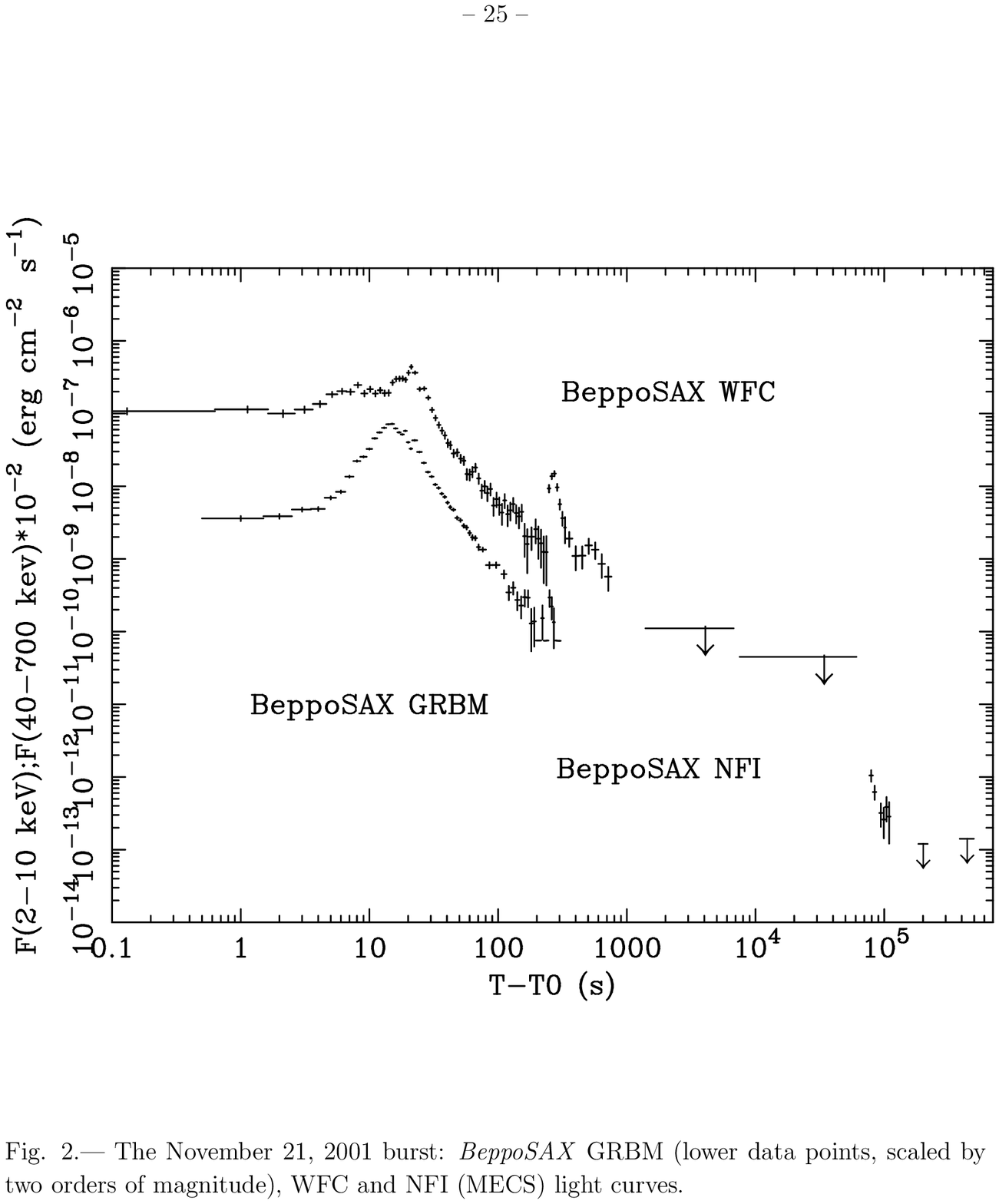}
   \caption{{\it Beppo-SAX} combined light curves of GRB 011121
     \citep[from][]{Piro05} from the Gamma-Ray Burst Monitor (GRBM),
     Wide Field Camera (WFC), and Narrow Field Instruments (NFI).}
   \label{fig:Piro}}
\hfill
     \parbox{2.4in}{
     \includegraphics[width=1.7in,angle=270]{figure2.ps}
   \caption{{\it Swift}/XRT light curve of GRB 060428A, which shows
     all of the phases seen in early GRB X-ray afterglows (but with an
     unusual dual-slope energy injection phase).}
   \label{fig:060428A}}
   \end{figure}

The typical {\it Beppo-SAX} afterglow observation began 6-8 hours after the
burst \citep{Frontera00}, and therefore early X-ray afterglows were
only observed for a handful of cases before 2004.  
\citet{Piro05} reported
{\it Beppo-SAX} light curves for two extremely bright bursts (GRB\,011121 and GRB\,011211) in which the early
afterglow was captured by the Wide Field Camera before the satellite was
re-pointed.  Both of these early afterglows contained bright X-ray
flares several hundred seconds after the burst, in which the flux
increased rapidly by a factor of several (e.g., the flare peaking at
$\sim 270$\3s in figure~\ref{fig:Piro}).
These events were attributed to the onset of the external shock in the
circumburst medium.
Such observations were quite rare until recently.

The {\it Swift} satellite \citep{Gehrels04} was specifically designed
to study early GRB afterglows by automatically and immediately slewing
to GRBs discovered on-board.  The satellite carries three instruments:
the wide-field Burst Alert Telescope \citep[BAT:][]{Barthelmy05}, which detects bursts and provides a
coarse position on the sky (typically $<3$ arcminutes); and two
narrow-field instruments: the X-ray
Telescope \citep[XRT:][]{Burrows05a} and the UV/Optical Telescope \citep[UVOT:][]{Roming05}.
Prompt slews are successfully executed for $> 80\%$ of bursts detected
by the BAT, and the afterglows are observed intensively by the XRT and
UVOT for hours, days or weeks, depending on the burst.
The XRT is designed to
switch automatically between different readout modes in order to
optimize the science as the afterglow fades \citep{Hill04}.
Because {\it Swift} is in a low Earth orbit (95 minute orbital period), there are gaps in
the light curves when the satellite is on the wrong side of the Earth
to view the burst.

In virtually every case, the XRT detects a fading X-ray
afterglow of the burst.  
The resulting data set has provided the first
comprehensive study of the early X-ray afterglows of GRBs in the time
interval of $\sim 0.1$ to $\sim 20$\3ks post-burst.  We have found that
the typical X-ray afterglow light curve follows a canonical template 
\citep{Nousek06,Bing06,OBrien06},
well illustrated by the case of GRB\,060428A
(figure~\ref{fig:060428A}).
This canonical light curve is a broken power law ($F_x \propto
[t-\rm{T}0]^{-\alpha_x}$, where T0 is the time at which the
burst was detected) with flares superimposed.
The first phase consists of a rapid decline in brightness from the burst itself 
(with $\alpha_x \sim 6$ in this case), which is
attributed to light delay effects from the expanding shock
\citep{Kumar00}.
This is followed by a flat plateau phase (in this case, the plateau
phase has two parts, with $\alpha_x \sim 0.1$ and 0.7), 
probably produced by energy injection into the
external shock; a `normal'
afterglow with $\alpha_x \sim 1.3$, and sometimes a late break to $\alpha_x
\sim 2.3$ when the edge of the collimated jet becomes visible.  
Detailed discussions of the physics behind these phases can be found
in \citet{Nousek06,Bing06}, and \citet{Panaitescu06}.
Although very few afterglows exhibit
all of these phases, most afterglows can be
characterized as a combination of at least two of them.

Roughly half of the afterglows have
X-ray flares superimposed on
this broken power-law light curve.  
The flare in GRB\,060428A is fairly weak, peaking only $\sim 30$\% in
excess of the underlying power law level, but is otherwise typical of
the flares observed with the XRT.
We note that in this case, the
flaring occurs during the plateau phase of the afterglow, that the
afterglow from the external forward shock is already in progress when the flare begins, and that the
light curve returns to this afterglow decay following the flare.
Several examples of these flares have been discussed 
by \citet{Burrows05b}, \citet{Romano06}, \citet{Falcone06}, and \citet{Pagani06}.
Here we provide a more comprehensive look at the properties of these
X-ray flares and what they can tell us about the GRB itself.


\section{Case Studies}

\subsection{Giant flares: GRBs 050502B and 060526}

\begin{figure}
       \centering
       \parbox{2.45in}{
         \includegraphics[width=2.3in]{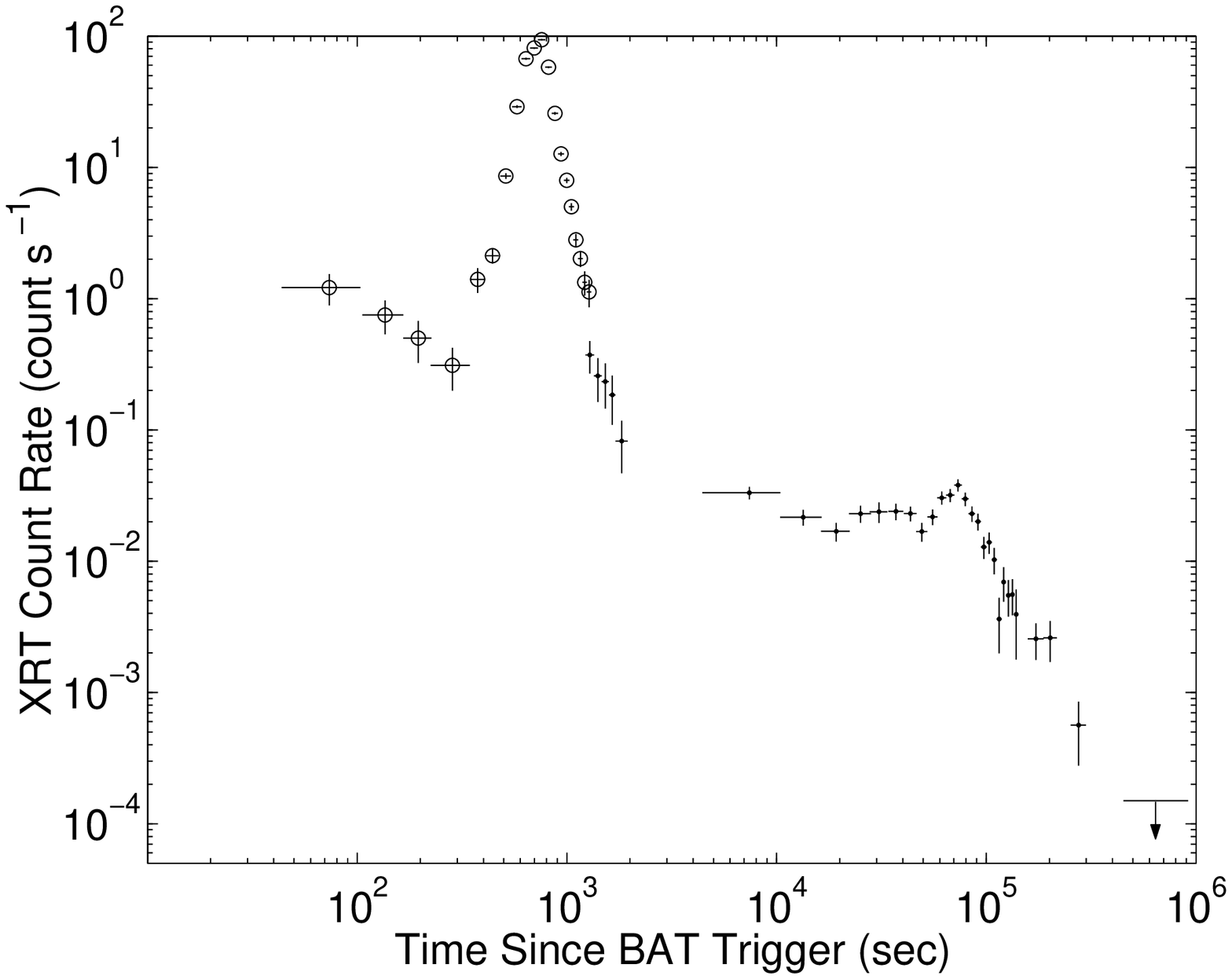}}
       \hfill
       \parbox{2.45in}{
         \includegraphics[width=2.8in]{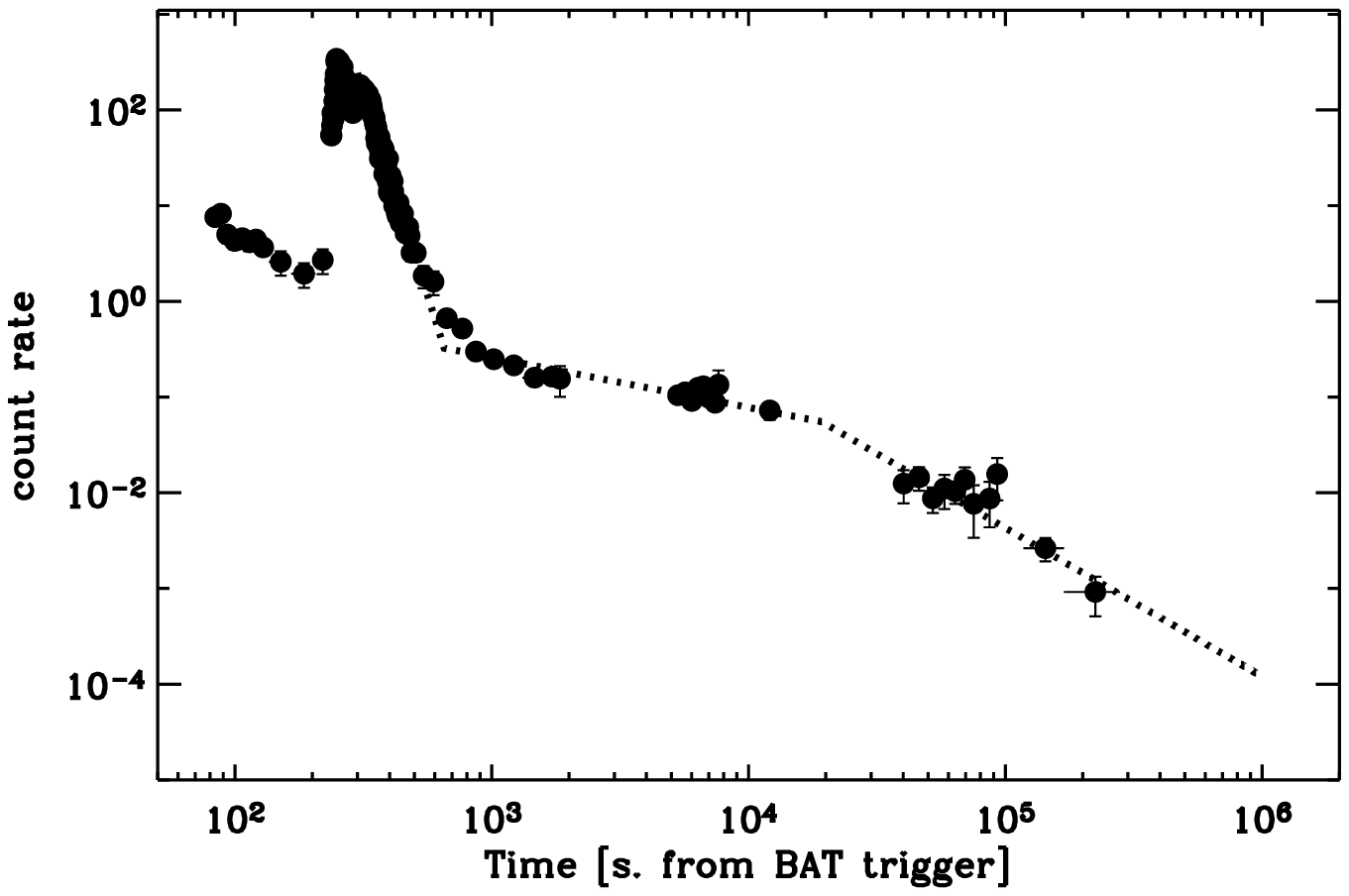}}
       \caption{Giant X-ray flares.  Left: X-ray light curve of
         GRB~050502B (0.2-10\3keV), from \citet{Falcone06}.
             Open circles are 
             Windowed Timing (WT) mode data and dots are
             Photon-Counting (PC) mode data (with pile-up corrections
             where necessary).
             Right: X-ray light curve of GRB~060526.  Dotted lines
             show the underlying afterglow.
             \label{fig:giant_flares}}
     \end{figure}

By way of contrast with the tiny flare of GRB\,060428A, we next
consider the giant flares seen in GRB\,050502B and  GRB\,060526 (figure~\ref{fig:giant_flares}).
GRB\,050502B is discussed in detail by \citet{Falcone06}.
The afterglow was initially faint, but the flux
increased dramatically by a factor of $\sim 500$, beginning about 345\3s
after the burst and peaking at about 700\3s.
The total X-ray fluence of the flare, $1 \times
10^{-6}$\3erg\3cm$^{-2}$ in the 0.2-10\3keV band, is comparable to 
the 15-350\3keV fluence of the gamma-ray burst itself.  
Like GRB\,060428A, the flare appears to be superimposed on a power law
decay that continues at the same slope after the flare ends; we will
refer to this as the `underlying afterglow' in subsequent discussion.
The underlying afterglow in the case of GRB\,050502B has a decay slope of
$\alpha_x \sim 0.8$.
The afterglow begins before the
commencement of XRT observations (and before the flare begins) and
continues after the flare ends.  The giant flare in
GRB\,060526 is remarkably similar, with a flux increase of about
$100\times$ relative to the underlying afterglow, which also has
$\alpha_x \sim 0.8$.  In this case, the giant flare clearly consists of
two separate, overlapping flares, peaking at about 220\3s and 300\3s.  The level of the underlying
afterglow before the flares is less clear than in GRB\,050502B.  
Giant flares like these so dominate
the light curve that they are reminiscent of the gamma-ray burst
itself, although at lower energies and much longer time-scales.  We
will see that these two differences (energy range and time-scale) may
be related.

\subsection{Flaring in a `naked' GRB: GRB\,050421}

\begin{figure}
          \centering
          \parbox{2.4in}{
            \includegraphics[width=2.3in]{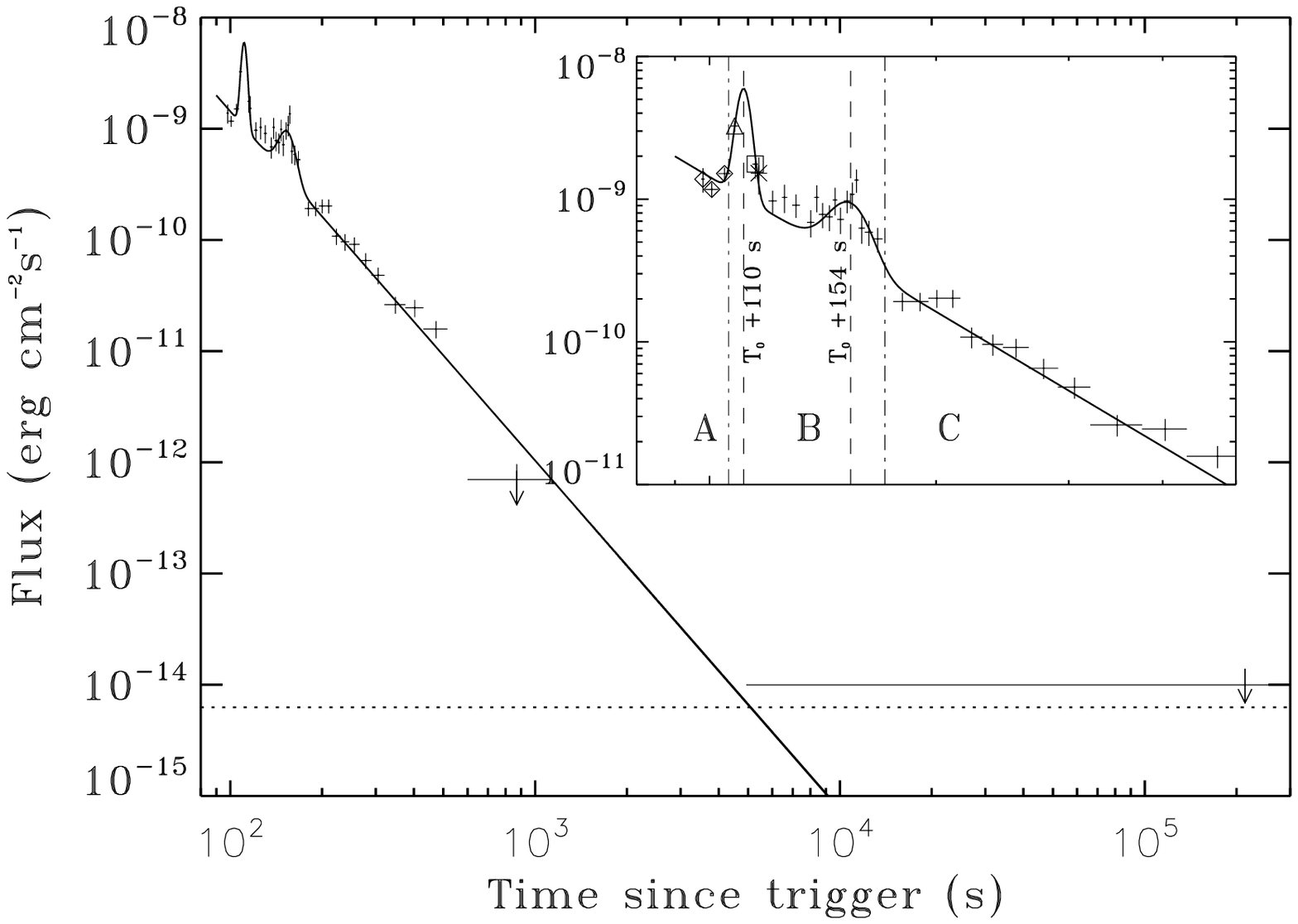}
            \caption{{\it Swift}/XRT 0.3-10\3keV light curve of GRB\,050421, a possible
              `naked' GRB with a strong flare \citep{Godet06}.}
            \label{fig:050421}}
           \hfill
           \parbox{2.4in}{
           \includegraphics[width=2.6in,clip,bb=5 335 550 700]{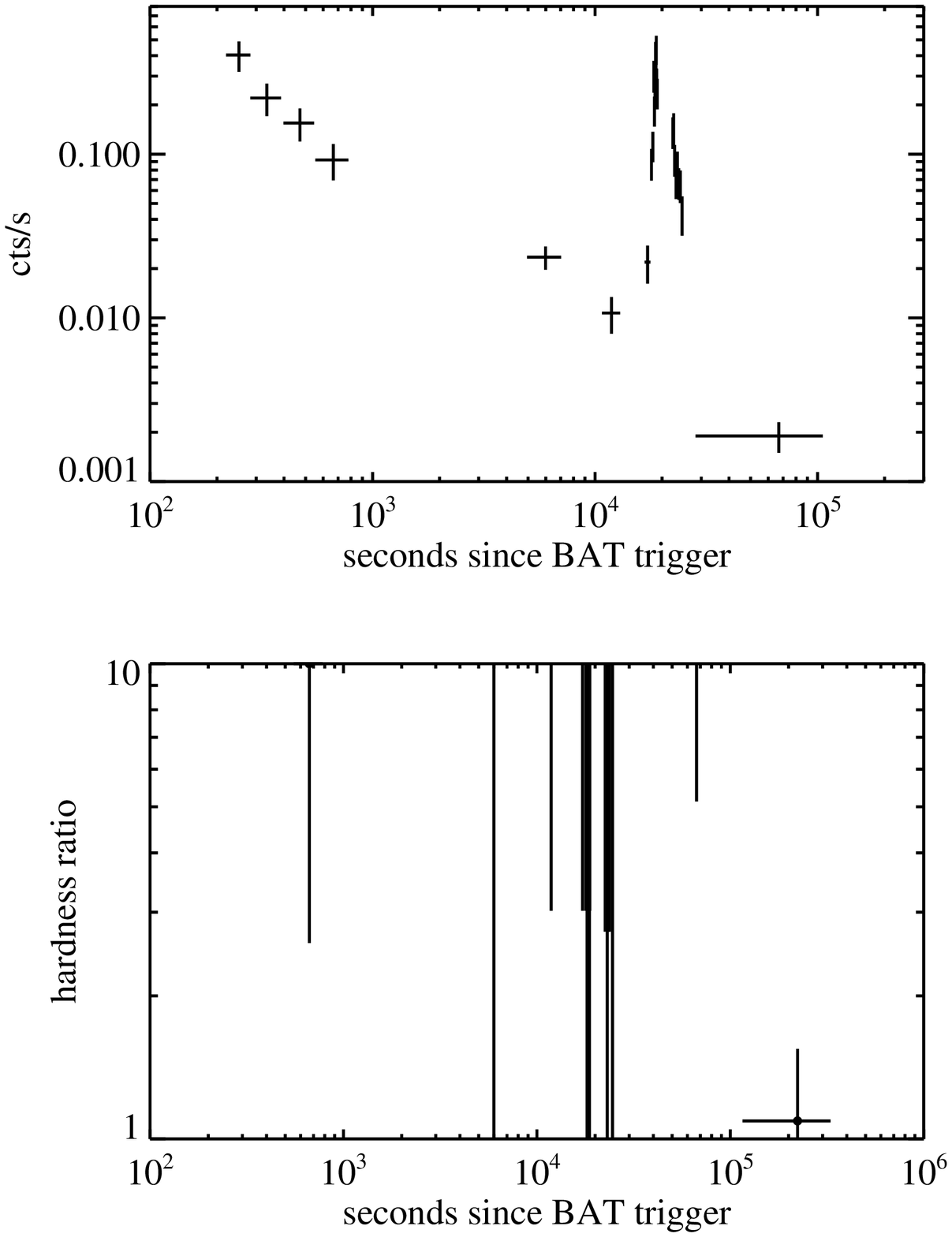}
           \caption{X-ray light curve of GRB\,050916.  A strong flare ($\sim$100
             $\times$ the afterglow brightness) at about 20\3ks is superimposed on an
             afterglow with decay index $\alpha_x \sim 1.5$.
             \label{fig:050916}}}
\vspace{-0.1in}
         \end{figure}

Figure~\ref{fig:050421} shows the X-ray light curve of GRB\,050421
\citep{Godet06}.
This light curve has a single power-law decay ($\alpha_x
\sim 3.1$) with at
least two flares superimposed, including the large (but poorly-sampled)
flare at T0+110\3s.  There is no hint in this light curve of any
flattening due to the presence of an afterglow from the forward shock;
in fact, there is no
evidence for any interaction of the shock with the external medium.
This GRB is therefore considered to be a possible `naked' GRB, i.e., a
GRB located in an extremely low density region \citep{Kumar00}.  The absence of
measured forward shock emission (to levels orders of magnitude below
typical afterglow fluxes), combined with the presence of multiple
flares in the light curve, therefore
provides strong evidence that these flares originate elsewhere,
perhaps in internal shocks within the relativistic outflow.

\subsection{Flares early and late}

X-ray flares or bumps have been observed in all phases of the X-ray
light curves, though they occur primarily in the first hour after the
burst.  
For example, the flares in GRB\,060607A occur during the initial steep decay,
whilst those in GRB\,050502B and GRB\,060526
(figure~\ref{fig:giant_flares}), as well as
GRB\,060428A (figure~\ref{fig:060428A}), occur during the flat plateau.
In a few cases we find a strong, rapidly rising and decaying flare
at quite late times,
such as the giant flare of GRB\,050916 (figure~\ref{fig:050916}).
This is clearly similar to the early strong flares seen in other
bursts.

GRB\,050730 (figure~\ref{fig:multiple_flares}) has a light curve that
continues to flare at times as late as 35\3ks post-burst.
We attribute this variability to flaring, even though distinct flares
are not visible in this case, because it is not possible to fit any simple
series of power laws to this light curve.  We note that
a power law decay slope is
established between 4.5\3ks and 7\3ks post-burst, and the light curve drops back to an extrapolation of
this decay slope in the interval around 25\3ks, consistent with
intervening flares
on top of an underlying afterglow.
By contrast, other mechanisms for increased count rates, such as refreshed shocks,
add energy into the external shock and thereafter the decay continues
at a higher level instead of dropping back to the original curve. 

\subsection{Flare multiplicity}
\begin{figure}
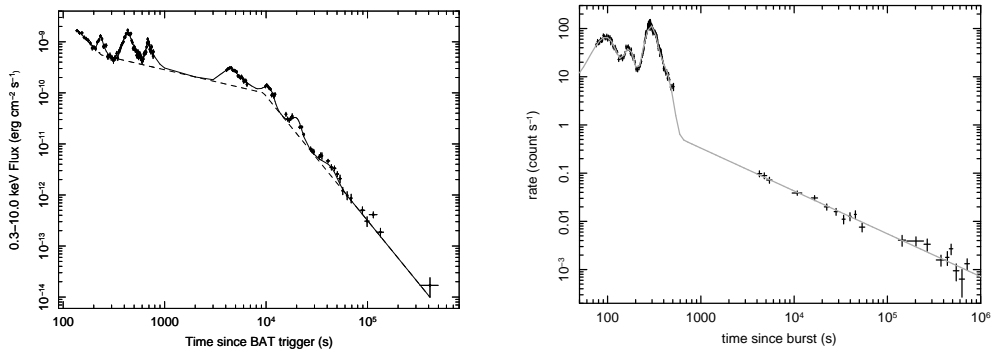

       \centering
       \parbox{2.45in}{
          \includegraphics[width=0.75\linewidth,angle=270]{figure6a.eps}}
       \hfill
       \parbox{2.45in}{
          \includegraphics[height=2.9in,angle=270,clip,bb=45 1 593 784]{figure6b.eps}}
       \caption{Left: X-ray light curve of
           GRB~050730 (0.3-10\3keV) \citep{Starling05,Pandey06}.
           The dashed line indicates a possible underlying power-law
           decay, and the solid line indicates a possible fit to the decay curve with a set of flares.  However, this light curve has so much variability that the
           level and slope of any underlying afterglow cannot be established with
           certainty.  At least 3 strong flares are seen in the WT mode
           data of the first orbit.
           Right: X-ray light curve of
           GRB~060111A.  Three strong flares are seen in the
           data from the first orbit.
             \label{fig:multiple_flares}}
     \end{figure}

In addition to isolated single flares, XRT light curves have many
examples of multiple flares, such as those shown in 
%
%
figure~\ref{fig:multiple_flares}. 
The light curves of GRB\,050730 and GRB\,060111A both begin with at least three X-ray flares of roughly
equal peak flux (or count rate).  
This argues against single-shot flare mechanisms, such as the onset of the
afterglow. 
\citep[][cite other evidence
  suggesting that the onset of the afterglow occurs very early in the
  light curve, probably prior to the rapid decline phase at the end of the last
  gamma-ray pulse.]{OBrien06}

\subsection{X-rays from GRB peaks}
\begin{figure}
       \centering
       \parbox{1.95in}{
         \includegraphics[width=1.4in,bb=116 179 373 493,clip]{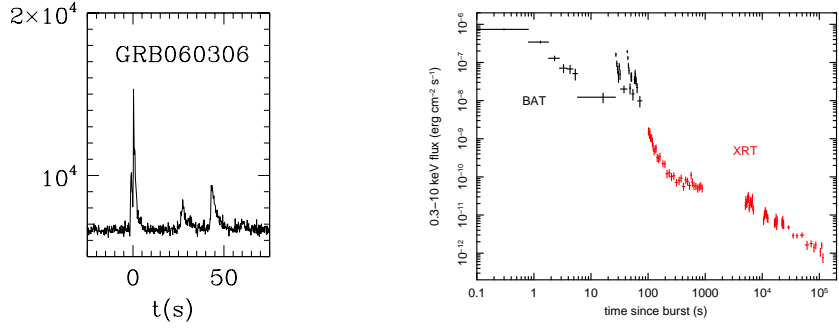}}
       \hfill
       \parbox{2.95in}{
         \includegraphics[height=2.6in,angle=270,clip,bb=45 1 593 784]{figure7b.eps}}
       \caption{Left: {\it Swift}/BAT light curve of GRB\,060306.
        The BAT measures three distinct peaks in the burst light
        curve, spread out over about 44\3s.  
        Right: Combined light curve of the BAT (extrapolated into the
        XRT energy band) and XRT observations of GRB\,060306.  No
        flares are seen in the XRT light curve, which begins with the
        tail of the last BAT peak.
             \label{fig:060306}}
\vspace{-0.15in}
     \end{figure}

Figure~\ref{fig:060306} shows the BAT and XRT light
curves for GRB\,060306.
In this burst, the BAT detected three distinct peaks of activity, with
the last occurring at about T0+44\3s.  By the time the XRT began
observations at T0+100\3s the prompt activity had ended, and the XRT
observations begin with the steep decay of the tail of the prompt emission.

\begin{figure}
       \centering
       \parbox{2.45in}{
           \includegraphics[width=1.9in,angle=270,clip,bb=45 1 593 784]{figure8.eps}
           \caption{BAT (black) and XRT (red) light curves of GRB\,060607A.
           The BAT data are extrapolated into the XRT band for direct
           comparison with the XRT fluxes.  There is good overlapping coverage of the final BAT peak at
           about T0+100\3s, with the XRT detecting several later,
           fainter flares.  Note that the XRT
           peak at 100\3s is much broader than the corresponding BAT peak, and that the
           broader soft X-ray profile is very similar to that of the
           later X-ray flares.
             \label{fig:BAT_XRT}}}
%
\hfill
          \parbox{2.45in}{
          \includegraphics[width=2.0in,clip,bb=59 80 530 670]{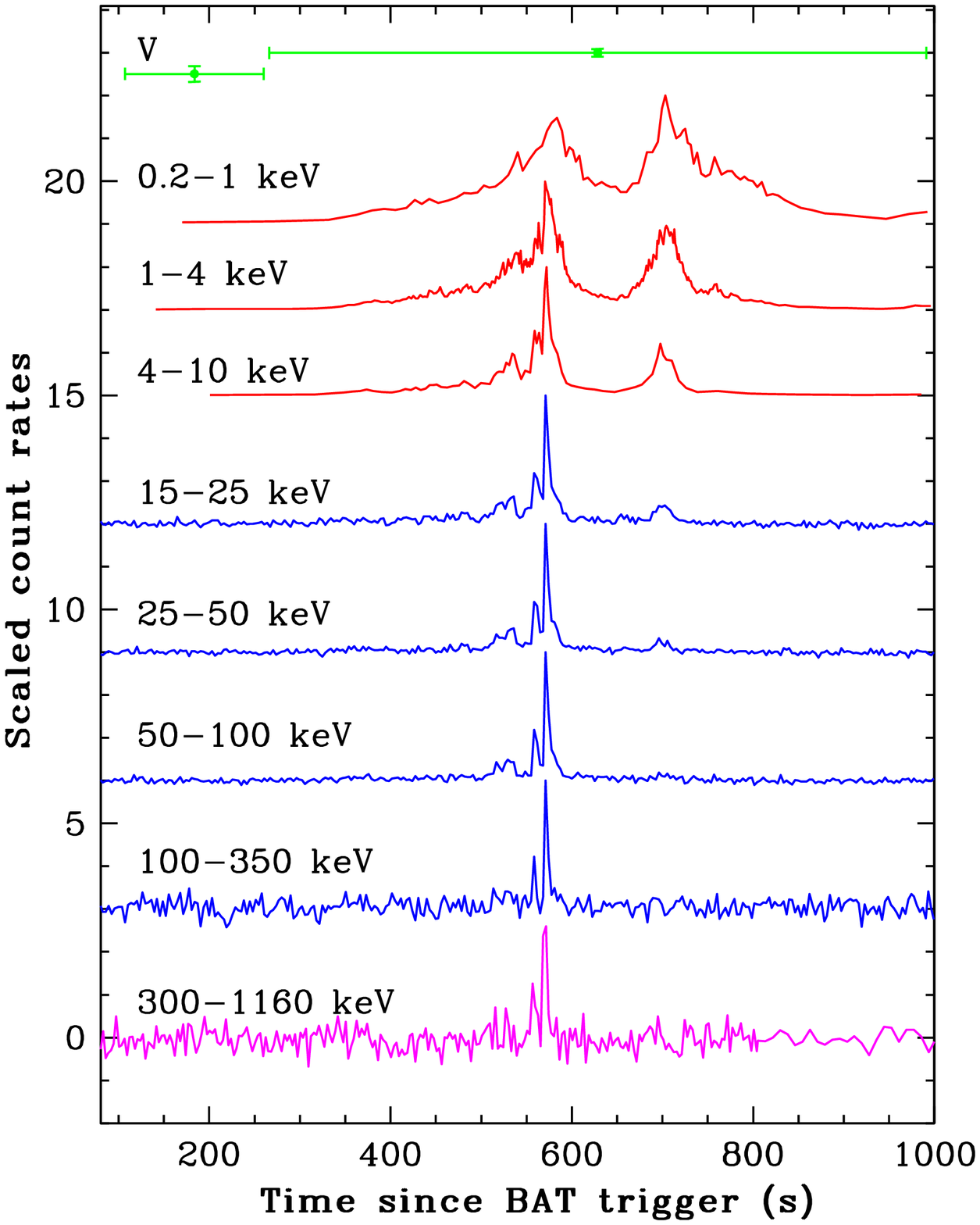}
          \caption{Light curves of the main burst of GRB\,060124 \citep[from][]{Romano06b} as measured by
            (from top to bottom): the {\it Swift} UVOT (V band), XRT (0.2-10\3keV), and
            BAT (15-350\3keV), and by Konus-Wind (300-1160\3keV).
            \label{fig:060124}}}
        \end{figure}

Had the XRT observations of GRB\,060306 started before T0+40\3s, both XRT and BAT
would have observed the third peak of the prompt emission.  In fact,
this has happened for a number of long-lasting bursts, including
GRB\,050117 \citep{Hill06}, GRB\,050713A \citep{Morris06},
GRB\,050820A \citep{Cenko06}, GRB\,060124 \citep{Romano06b},
GRB\,060418, 
GRB\,060510B,
GRB\,060607A, and GRB\,060714.
In figure~\ref{fig:BAT_XRT} we show a recent example.
The BAT and XRT observed this burst in the 15-150\3keV and 
0.2-10\3keV bands, respectively.  Spectral evolution is evident
in the different relative strengths and widths of the peaks as
detected by the two instruments.  In general, the soft X-ray
manifestations of the prompt peaks are significantly broader than the
peaks observed in the hard X-ray band, as is clear from the light
curves of GRB\,060607A, in which it
is also clear that the soft X-ray peak is very similar to the later
X-ray flares.  The energy dependence of the prompt peaks is
particularly evident in GRB\,060124 \citep{Romano06b}, which was
detected by both {\it Swift} and Konus-Wind and so has very broad-band
coverage.
Figure~\ref{fig:060124} shows the main burst of GRB\,060124 as a function of energy
from the optical (V band) to $\gamma$-ray (1\3MeV).  This burst had a particularly long period of
activity, with a precursor that triggered the BAT nearly 10 minutes
before the main burst.  This later pulse is very narrow
at high energies, but is quite broad in the soft X-ray band, and is
followed by a very similar, but much softer X-ray flare that is barely
detected by the BAT above 15\3keV.
The only distinction between this later X-ray flare and the earlier
gamma-ray burst appears to be the spectral distribution of the
photons.
We note that this event would have been classified as an X-Ray Flash
(XRF) had it consisted only of this later flare.

\subsection{Spectral evolution during flares}

These giant flares are so bright that detailed spectral analysis is
possible.  Detailed spectral analysis for GRB\,050502B is discussed in
\citet{Falcone06}, who also showed that a simple hardness ratio shows
strong spectral evolution, increasing  
%
%
steeply at the beginning of the flare,
peaking well before the flare itself peaks, and then dropping
gradually through the remainder of the flare, finally returning to the
same level established in the afterglow emission before the flare.
This behaviour is also seen in other bursts with sufficient statistics:
e.g., GRBs\,050607 \citep{Pagani06}, 050822 \citep{Godet06}, 051117A, and 060607A \citep{Guido06}.
This spectral evolution results from a
hard-to-soft transition in the X-ray flare similar to that commonly seen in GRB prompt
emission \citep[e.g.,][]{Ford95,Frontera00,Piro05}.

\begin{figure}
      \centering
      \parbox{2.45in}{
      \includegraphics[height=0.65\linewidth,angle=0,clip,bb=100 70 990 530]{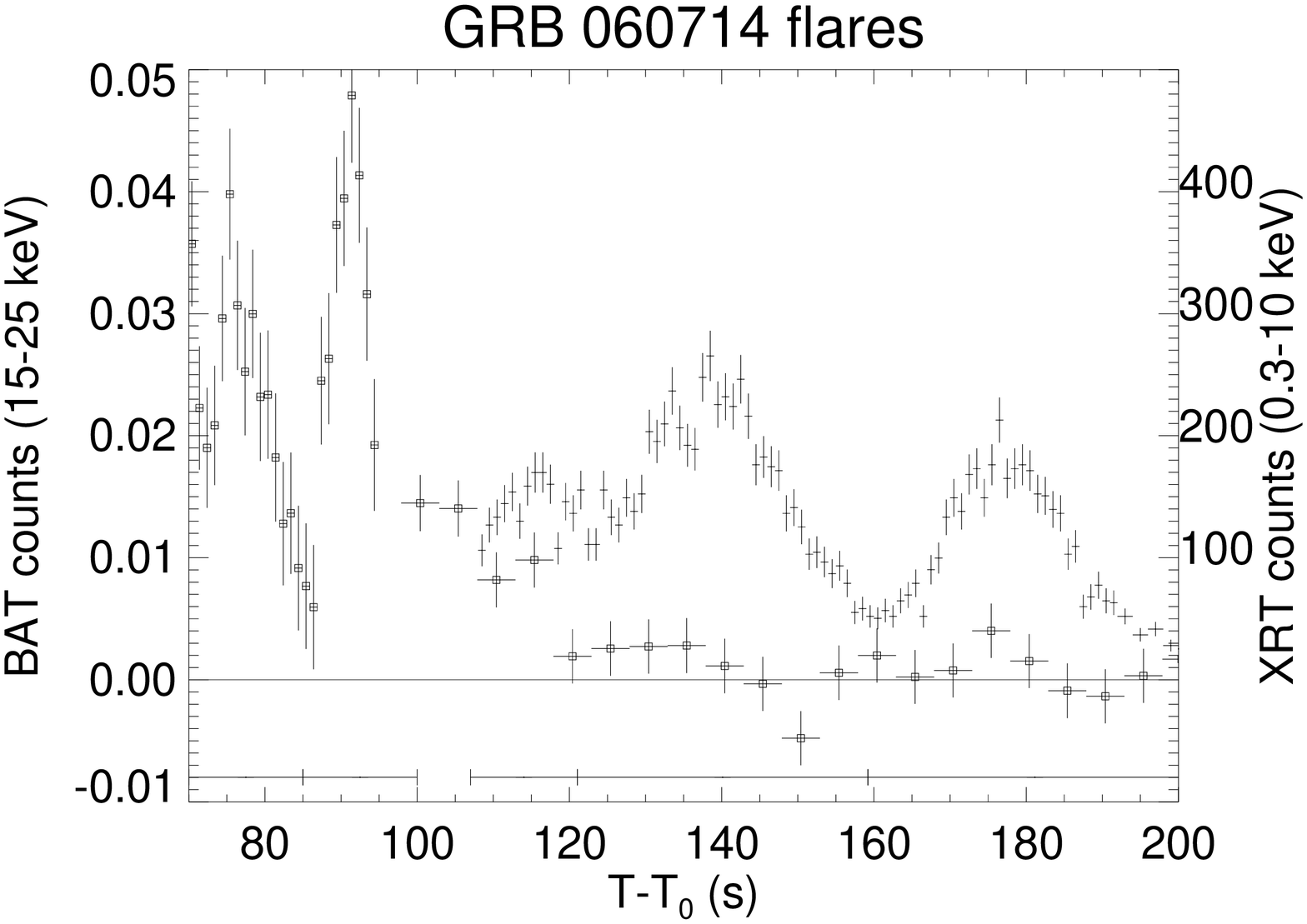}}
      \hfill
      \parbox{2.45in}{
        \includegraphics[height=0.65\linewidth,angle=0,clip,bb=100 70 970 530]{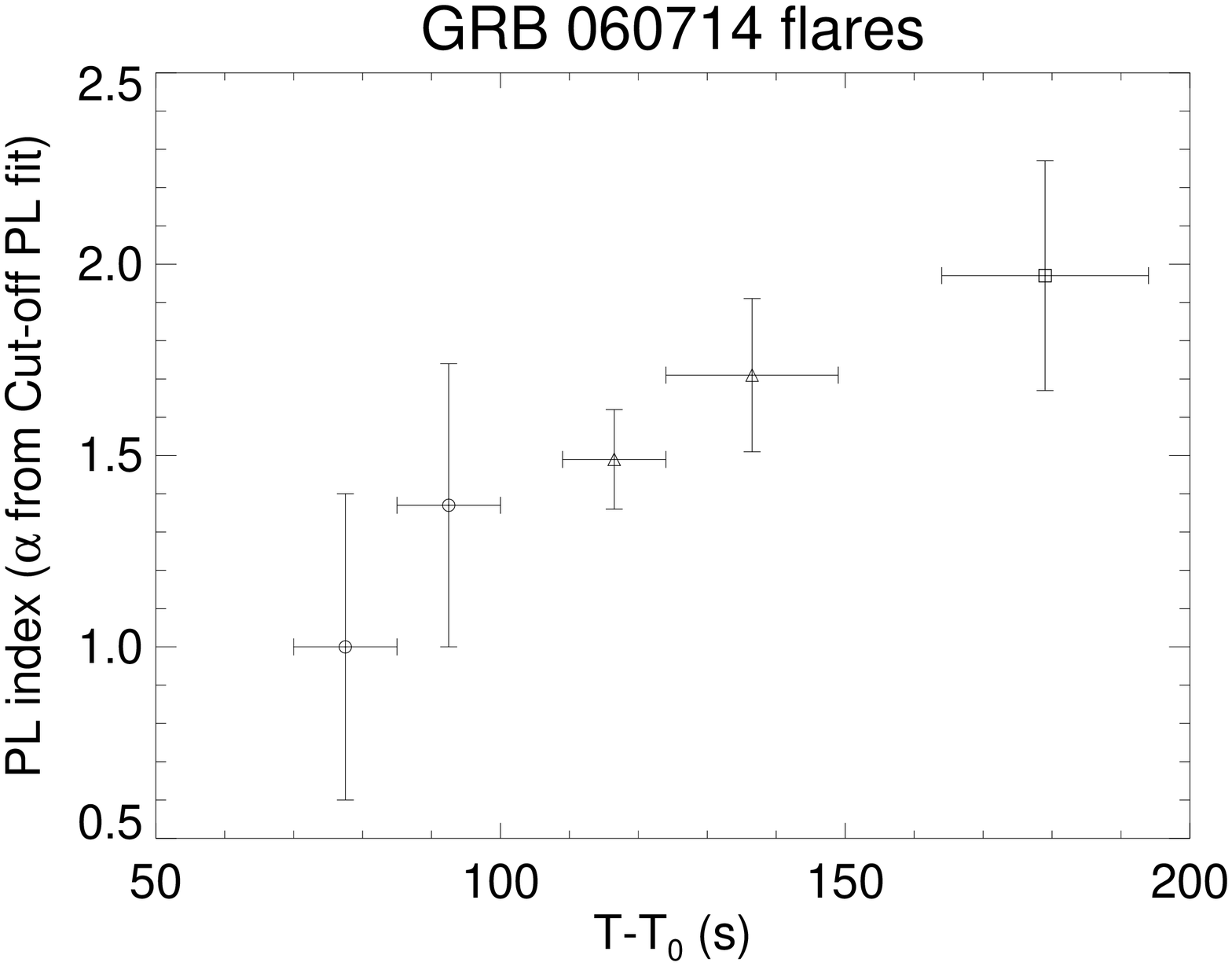}}
      \caption{Left: BAT (boxes) and XRT (crosses) light curves for GRB\,060714.  The
        horizontal bars indicate the time periods used in the spectral
        fits of the individual BAT and XRT peaks.
        Right: Photon index for a power law spectral model with
        exponential cut-off at high energies for five peaks in the
        BAT/XRT light curve of GRB\,060714.  The first two peaks were
        fit to only the BAT data, the third peak was fit to XRT and
        BAT data simultaneously, and the last two peaks were fit to
        only the XRT data.
         \label{fig:060714}}
\end{figure}

A particularly interesting example is GRB\,060714, in which a series
of pulses in the BAT energy range (15-350\3keV) is followed by a
similar series in the XRT band (0.2-10\3keV).  The spectra of these
pulses can be fit with a cut-off power law model, giving clear
evidence of spectral evolution (figure~\ref{fig:060714}),
with the spectrum becoming successively softer with each new peak.

Similar behavior is observed in a number of other flares, and 
includes apparent transitions of the emission from the gamma-ray or hard X-ray
bands into the soft X-ray band as time progresses.  For example,
observationally we see in both GRB\,060714 and 
GRB\,060418 
that the final, weak BAT pulse is 
seen in the XRT as a strong flare.  
This transition to softer pulses can be clearly seen in GRB\,060124 (figure~\ref{fig:060124}).
As a final example, our analysis of GRB\,050416A finds evidence
that $E_{peak}$ moves from the BAT band into the XRT band
\citep{Mangano06}.
This suggests a strong connection between the X-ray flares and the
prompt emission - they appear to be the same phenomenon, simply
expressed at lower energies at later times.

\subsection{Flares in short GRB afterglows: GRBs 050724 and 051210}

So far we have only discussed flares in long GRBs, which are thought
to be the result of the collapse of massive stars into black holes.
Two short GRBs, GRB\,050724
\citep{Barthelmy05b,Campana06,Grupe06}
and GRB\,051210 \citep{LaParola06}, also have flares in their X-ray light
curves.  
This requires that short GRB models, which typically invoke mergers of
compact objects (two neutron stars or a neutron star and a black
hole),
must also be capable of producing flares.
Of particular interest in this regard is the very late and energetic event at about
12 hours post-burst in GRB\,050724, a very late time-scale for a merger
process thought to last only milliseconds.

\section{Statistical analysis}

Having examined some `case studies' illustrative of the properties of
the X-ray flares, we now turn to a more objective statistical
treatment of their properties.
We are undertaking a comprehensive study of the properties
of flares using a set of 77 flares in 33 XRT
afterglows.  Here we simply highlight a few preliminary results 
related to the temporal properties of X-ray flares in
GRB afterglows.

\begin{figure}
       \centering
      \parbox{2.45in}{
       \includegraphics[width=0.85\linewidth]{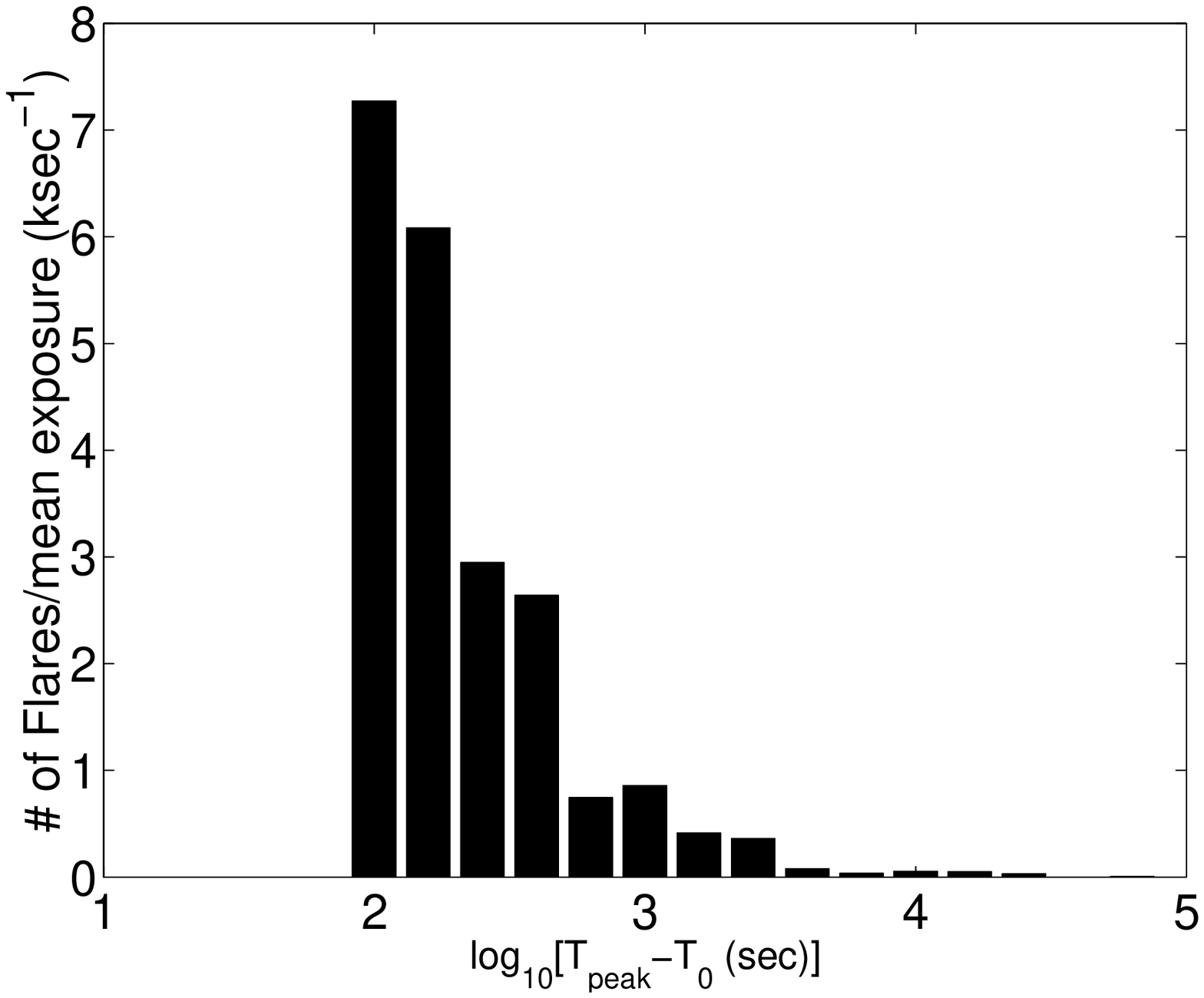}
       \caption{Distribution of peak times of X-ray flares in our
         sample, normalized to the exposure time in each bin.
         \label{fig:Tpeak}}}
     \parbox{2.45in}{
       \includegraphics[width=2.2in]{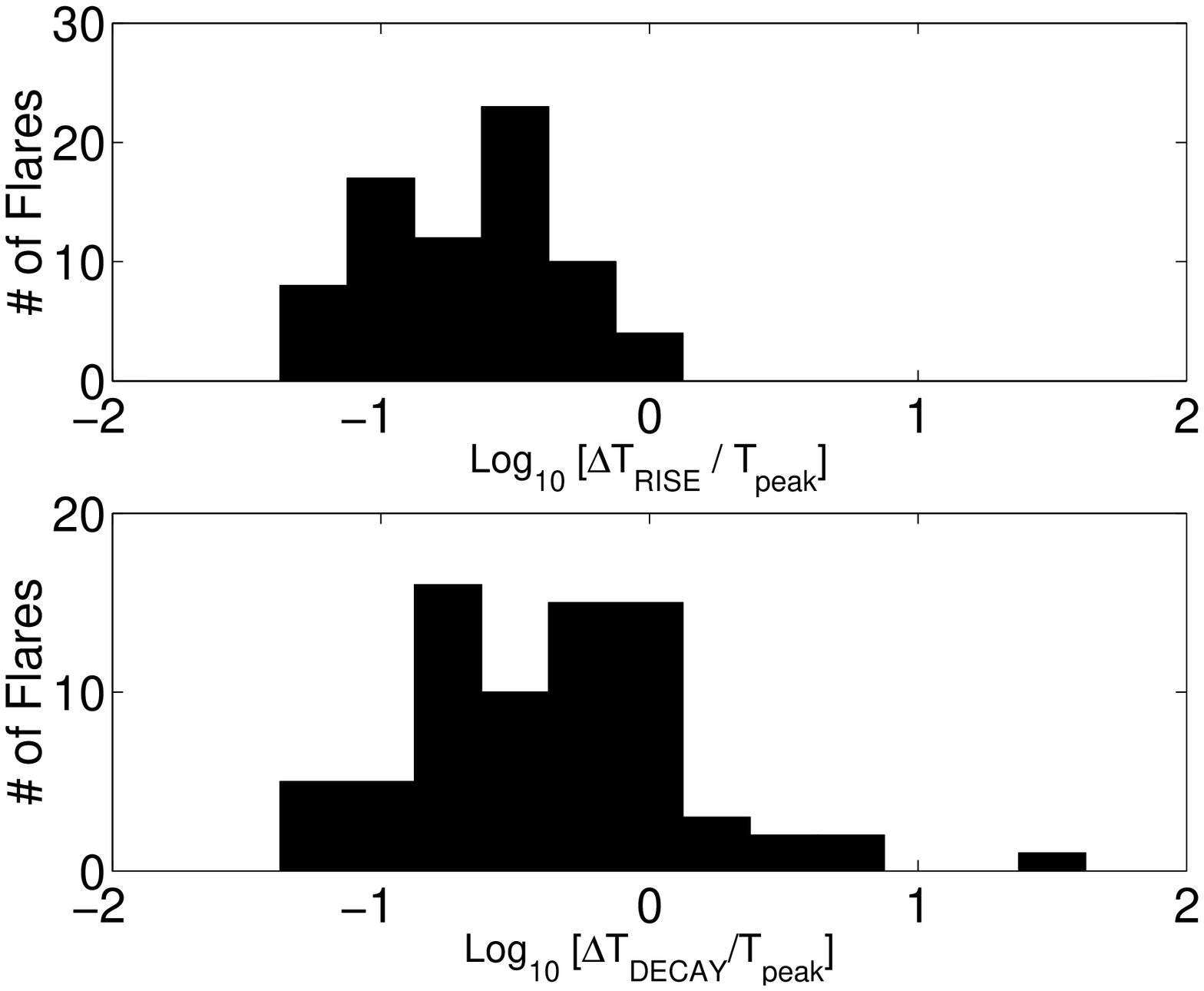}
       \caption{Distributions of rise and decay times for flares in our sample.
         \label{fig:dt}}}
     \end{figure}

The distribution of peak times of flares is shown in
figure~\ref{fig:Tpeak}.  The times are in the observer frame,
uncorrected for redshift (since we do not have redshifts for many of
these bursts).
The distribution follows a power law with slope of -1.1, with flares strongly concentrated at
early times.  This is not surprising if the X-ray flares represent the
last sputters of the GRB, but further work must be done to determine whether this
distribution can be used to rule out other models for the flares.


An important clue to the nature of the flares can be found by analysing their
rise and decay times.  Mechanisms such as external shocks encountering
clouds in the circumburst medium are expected to produce rather slow
rise and decay times, with decay times characterized by $\Delta t/t \sim
1$ \citep{Ioka05,Nakar06}.  By contrast, the pulses associated with collisions of shocks
within the relativistic outflow (`internal shocks') are typically much
faster, with $\Delta t/t << 1$.  The X-ray flares span both ranges,
with some having $\Delta t/t \sim 0.1$ or less.  
For example, figure~\ref{fig:050916} shows a case of a narrow, late
flare.

In
order to ascertain the rise and decay times, we have fit each flare in our sample of 77
flares with a model consisting of an underlying power law decay with a
flare superimposed on it.  We fit both the rising portion of
the flare and its decay with power laws.  We then define $\Delta t_{rise}$ as
$t_p - t_1$, where $t_p$ is the peak time (the time when the
rising power law intersects the decaying power law), and $t_1$ is the
time when the rising power law intersects the underlying afterglow's
power law decay.
Similarly, the decay time is defined as $\Delta t_{decay} = t_2-t_p$,
where $t_2$ is the time when the falling power law tail of the flare
crosses the underlying power law decay.
Figure~\ref{fig:dt} shows the distributions of
rise and decay times for our sample.
The distributions peak near $\Delta t/t \sim 0.3$, 
uncomfortably short for external shocks.

\section{Discussion}

X-ray flares in GRB afterglows could be produced by a variety of
mechanisms, including internal shocks (the same mechanism that
produces the GRB prompt emission) at late times, external forward
shocks (either the onset of the forward shock or encounters by the
forward shock of density gradients or clouds), or external reverse shocks.
Based on the observations described above, we believe that there is
strong evidence that X-ray flares in early GRB afterglows are not
produced by the external shock (forward or reverse).
This view is supported by:
1) the rapid rise and decay times of most X-ray flares;
2) the multiplicity of many X-ray flares, which cannot be explained
      by `single-shot' mechanisms like the onset of the afterglow;
3) the fact that some flares are clearly superimposed on a
      pre-existing afterglow decay;
4) the observation of flares in `naked' afterglows with no evidence for
      forward shock emission (e.g., GRB\,050421);
5) the enormous increase in flux in giant flares.

By contrast, all of these characteristics are easily accommodated by
internal shocks.  Internal shocks can also explain the following
observational characteristics of X-ray flares:
1) the similarity between the shapes of X-ray flares and
      X-ray peaks in the prompt emission;
2) the spectral evolution of X-ray flares, both within individual
      flares and between successive flares, which mimics that commonly
      seen in the prompt emission.
Additional support for the view that X-ray flares are due to internal
shocks comes from the analysis of \citet{Liang06}, who showed that the
decay slopes of X-ray flares are consistent with the expectations of
the `curvature effect' \citep[time delay effects from an extended shock
front following the cessation of the emission;][]{Kumar00}
We conclude that X-ray flares in GRB afterglows are probably the result of
late-time internal shocks.


These late internal shocks can be produced in at least two ways.  
In the internal shock model \citep{Rees_Mes94}, the central engine activity produces a series of
shells of outflowing material with a variety of Lorentz factors, and the GRB
prompt emission results from `internal shocks' as these shells collide in the
relativistic outflow.  Such collisions can occur over a wide range of
timescales, with late shocks resulting from collisions between shells
with Lorentz factors that differ by only a small amount.
These could produce X-ray flares at relatively late times.
The fact that these late-time collisions must occur at large radii may
lead to a natural explanation for the relatively soft photon emission.
On the other hand, this mechanism may not be able to produce the
required efficiency of conversion of kinetic energy of the jet to
afterglow flux, due to the weakness of the resulting shock \citep{Zhang06}.

A second possibility is that the late internal shocks are the result
of continued activity in the central engine itself.  In this case, the
central engine must remain active for time-scales up to ten times
longer than the prompt emission, with important implications for GRB
models.  For long GRBs, the leading collapsar model would
need to explain in-fall of material into the central black
hole over time-scales of hours.  For short GRBs, a similar problem
exists in the context of merging compact objects.
We note that the redshift of the GRB can play a role in delaying
flaring activity in the observer's frame, and also in broadening the
flares.  This cannot be the sole explanation for late, broad X-ray flares, however; even in the
case of GRB\,050904, with redshift $z=6.29$, flaring occurs for over an
hour in the burst rest frame \citep{Watson06,Cusumano06}.
Furthermore, the mean redshift of flaring GRBs ($z=2.6$ for 14 flaring
GRBs with known redshifts) does not differ significantly from those
without flares, so the X-ray flares are not simply due to prompt
emission delayed, broadened, and reduced in peak energy by high redshifts.
We conclude that the central engines of GRBs remain active for hours
following the GRB event.

%
%
%

\begin{acknowledgements}
This work is supported at Penn State by NASA contract NAS5-00136, at OAB by ASI grant I/R/039/04, 
and at the University of Leicester by funding from PPARC.
\end{acknowledgements}

\label{lastpage}


\begin{thebibliography}{}
\bibitem[Barthelmy et al.(2005$a$)]{Barthelmy05} Barthelmy, S. D., et al.
      2005$a$ The Burst Alert Telescope (BAT) on the SWIFT Midex Mission. {\it Space Sci. Rev.}, {\bf 120}, 143-164.

\bibitem[Barthelmy et al.(2005$b$)]{Barthelmy05b} Barthelmy, S. D., et al.
 2005$b$ An origin for short $\gamma$-ray bursts unassociated with current star formation. \nature, {\bf 438}, 994-996.

\bibitem[Burrows et al.(2005$a$)]{Burrows05a} Burrows, D. N., et al.\
      2005$a$ The Swift X-Ray Telescope.
      {\it Space Sci. Rev.}, {\bf 120}, 165-195.

\bibitem[Burrows et al.(2005$b$)]{Burrows05b} Burrows, D. N. et
      al. 2005$b$ Bright X-ray Flares in Gamma-Ray Burst Afterglows. {\it Science}, {\bf 309}, 1833-1835.

\bibitem[Campana et al.(2006)]{Campana06} Campana, S., et al. 2006 
      The X-ray afterglow of the short gamma ray burst 050724.
      \aap, {\bf 454}, 113-117.

\bibitem[Cenko et al.(2006)]{Cenko06} Cenko, S. B., et al. 2006,
      \apj, in press, {\it astro-ph/0608183}.

\bibitem[Chincarini(2006)]{Guido06} Chincarini, G. 2006 GRBs with the Swift satellite. 
In {\it Proc. of Vulcano Workshop 2006, Frontier Objects in Astrophysics
  and Particle Physics} (ed. 
F. Giovannelli \& G. Mannocchi), in press.  Bologna:Italian Physical Society, 
Editrice Compositori, {\it astro-ph/0608414}.

\bibitem[Costa et al.(1997)]{Costa97} Costa, E., et al.\ 1997 
      Discovery of an X-ray afterglow associated with the gamma-ray burst of 28 February 1997.
      \nat, {\bf 387}, 783-785.

\bibitem[Cusumano et al.(2006)]{Cusumano06} Cusumano, G., et al. 2006 
      Gamma-ray bursts: Huge explosion in the early Universe. 
      \nature, {\bf 440}, 164.


\bibitem[Falcone et al.(2006)]{Falcone06} Falcone, A., et al. 2006
      The Giant X-Ray Flare of GRB 050502B: Evidence for Late-Time Internal Engine Activity.
      \apj, {\bf 641}, 1010-1017.

\bibitem[Ford et al.(1995)]{Ford95} Ford, L. A., et al. 1995
      BATSE observations of gamma-ray burst spectra. 2: Peak energy evolution in bright, long bursts.
      \apj, {\bf 439}, 307-321.

\bibitem[Frontera et al.(2000)]{Frontera00} Frontera, F., et al. 2000
      Prompt and Delayed Emission Properties of Gamma-Ray Bursts Observed with BeppoSAX.
      \apjs, {\bf 127}, 59-78

\bibitem[Gehrels et al.(2004)]{Gehrels04} Gehrels, N., et al. 2004
      The Swift Gamma-Ray Burst Mission.
      \apj, {\bf 611}, 1005-1020.

\bibitem[Godet et al.(2006$a$)]{Godet06} Godet, O., et al. 2006$a$
      X-ray flares in the early Swift observations of the possible naked gamma-ray burst 050421.
      \aap, {\bf 452}, 819-825.

\bibitem[Godet et al.(2006$b$)]{Godet06b} Godet, O., et al. 2006$b$
      Swift XRF 050822: Internal processes versus cocoon breakout.
      \aap, submitted.

\bibitem[Grupe et al.(2006)]{Grupe06} Grupe, D., Burrows, D. N.,
      Patel, S. K., Kouveliotou, C., Zhang, B., \peterm, P., Wijers,
      R. A. M., \& Gehrels, N. 2006
      Jet breaks in Short Gamma-ray Bursts. I: The Uncollimated
      Afterglow of GRB 050724.
      \apj, in press, {\it astro-ph/0603773}.

\bibitem[Hill et al.(2004)]{Hill04} Hill, J.E., et al. 2004
      Readout modes and automated operation of the Swift X-ray Telescope.
      {\it Proc. SPIE}, {\bf 5165}, 217-231.

\bibitem[Hill et al.(2006)]{Hill06} Hill, J.E., et al. 2006
      GRB 050117: Simultaneous Gamma-Ray and X-Ray Observations with the Swift Satellite.
      \apj, {\bf 639}, 303-310.

\bibitem[Ioka et al.(2005)]{Ioka05} Ioka, K., Kobayashi,
      S., \& Zhang, B. 2005
      Variabilities of Gamma-Ray Burst Afterglows: Long-acting Engine, Anisotropic Jet, or Many Fluctuating Regions?
      \apj, {\bf 631}, 429-434.

\bibitem[Kumar \& Panaitescu(2000)]{Kumar00} Kumar, P., \&
      Panaitescu, A. 2000
      Afterglow Emission from Naked Gamma-Ray Bursts.
      \apj, 541, L51-L54.

\bibitem[La Parola et al.(2006)]{LaParola06} La Parola, V., et
      al. 2006
      GRB 051210: Swift detection of a short gamma ray burst.
      \aap, {\bf 454}, 753-757.


\bibitem[Liang et al. (2006)]{Liang06} Liang, E., et al. 2006
      Testing the Curvature Effect and Internal Origin of Gamma-Ray Burst Prompt Emissions and X-Ray Flares with Swift Data.
      \apj, {\bf 646}, 351-357.

\bibitem[Mangano et al.(2006)]{Mangano06} Mangano, V., et al. 2006
      Swift XRT Observations of the Afterglow of XRF 050416A.
      \apj, in press, {\it astro-ph/0603738}.

\bibitem[M\'{e}sz\'{a}ros(2002)]{Meszaros02} M\'{e}sz\'{a}ros, P. 2002
      Theories of Gamma-Ray Bursts.
      {\it ARA\&A}, {\bf 40}, 137

\bibitem[Morris et al.(2006)]{Morris06} Morris, D. C., et al. 2006
      GRB 050713A: High Energy Observations of the GRB Prompt and Afterglow Emission.
      \apj, in press, {\it astro-ph/0602490} 

\bibitem[Nakar \& Granot(2006)]{Nakar06} Nakar, E., \& Granot,
      J. 2006
      Smooth Light Curves from a Bumpy Ride: Relativistic BlastWave Encounters a Density Jump.
      \mnras, submitted, {\it astro-ph/0606011}

\bibitem[Nousek et al.(2006)]{Nousek06} Nousek, J. A., et al. 2006
      Evidence for a Canonical Gamma-Ray Burst Afterglow Light Curve in the Swift XRT Data.
      \apj, {\bf 642}, 389-400.

\bibitem[O'Brien et al.(2006)]{OBrien06} O'Brien, P. T., et al. 2006
      The Early X-Ray Emission from GRBs.
      \apj, {\bf 647}, 1213-1237.

\bibitem[Pagani et al.(2006)]{Pagani06} Pagani, C., et al. 2006
      The Swift X-Ray Flaring Afterglow of GRB 050607.
      \apj, {\bf 645}, 1315-1322.

\bibitem[Panaitescu et al.(2006)]{Panaitescu06} Panaitescu, A.,
      \meszaros, P., Gehrels, N., Burrows, D., \& Nousek, J. 2006
      Analysis of the X-ray emission of nine Swift afterglows.
      \mnras, {\bf 366}, 1357-1366.

\bibitem[Pandey et al.(2006)]{Pandey06} Pandey, S. B., et al. 2006
      Multi-wavelength afterglow observations of the high redshift GRB 050730.
      \aap, in press, {\it astro-ph/0607471}.

\bibitem[Piro et al.(2005)]{Piro05} Piro, L., et al. 2005
      Probing the environment in Gamma-ray bursts: the case of an
      X-ray precursor, afterglow late onset and wind vs constant
      density profile in GRB 011121 and GRB 011211.
      \apj, {\bf 623}, 314-324.

\bibitem[Rees \& \meszaros (1994)]{Rees_Mes94} Rees, M. J., \&
      \meszaros, P. 1994
      Unsteady outflow models for cosmological gamma-ray bursts.
      \apj, {\bf 430}, L93-L96.

\bibitem[Romano et al.(2006$a$)]{Romano06} Romano, P. et al. 2006$a$
      X-ray flare in XRF 050406: evidence for prolonged engine activity.
      \aap, {\bf 450}, 59-68.

\bibitem[Romano et al.(2006$b$)]{Romano06b} Romano, P., et al. 2006$b$
      Panchromatic study of GRB 060124: from precursor to afterglow.
      \aap, {\bf 456}, 917-927.

\bibitem[Roming et al.(2005)]{Roming05} Roming, P.W.A., et al., 2005
      The Swift Ultra-Violet/Optical Telescope.
      {\it Space Sci. Rev.}, {\bf 120}, 95-142.

\bibitem[Starling et al.(2005)]{Starling05} Starling, R.L.C., et
      al. 2005
      Gas and dust properties in the afterglow spectra of GRB 050730.
      \aap, {\bf 442}, L21-L24.

\bibitem[Watson et al.(2006)]{Watson06} Watson, D., Reeves, J. N.,
      Hjorth, J., Fynbo, J. P. U., Jakobsson, P., Pedersen, K.,
      Sollerman, J., Castro Cer\'{o}n, J. M., McBreen, S., Foley,
      S. 2006
      Outshining the Quasars at Reionization: The X-Ray Spectrum and
      Light Curve of the Redshift 6.29 Gamma-Ray Burst GRB 050904.
      \apj, {\bf 637}, L69-L72.

\bibitem[Zhang et al.(2006)]{Bing06} Zhang, B., Fan, Y.Z., Dyks, J., 
      Kobayashi, S., \meszaros, P., Burrows, B.N., Nousek, J.A., \& 
      Gehrels, N., 2006
      Physical Processes Shaping Gamma-Ray Burst X-Ray Afterglow Light
      Curves: Theoretical Implications from the Swift X-Ray Telescope Observations.
      \apj, {\bf 642}, 354-370.

\bibitem[Zhang(2006)]{Zhang06} Zhang, B., 2006 Gamma-Ray Burst
      Afterglows.  {\it Advances in Space Research}, submitted.

\end{thebibliography}
\end{document}